# Continuous Mott transition in semiconductor moiré superlattices

Tingxin Li[1], Shengwei Jiang[2], Lizhong Li[1], Yang Zhang[3], Kaifei Kang[1], Jiacheng Zhu[1], Kenji Watanabe[4], Takashi Taniguchi[4], Debanjan Chowdhury[2], Liang Fu[3], Jie Shan[1,2,5*], Kin Fai Mak[1,2,5*]

[1]School of Applied and Engineering Physics, Cornell University, Ithaca, NY, USA
[2]Laboratory of Atomic and Solid State Physics, Cornell University, Ithaca, NY, USA
[3]Department of Physics, Massachusetts Institute of Technology, Cambridge, MA, USA
[4]National Institute for Materials Science, 1-1 Namiki, Tsukuba, Japan
[5]Kavli Institute at Cornell for Nanoscale Science, Ithaca, NY, USA

Correspondence to: jie.shan@cornell.edu, kinfai.mak@cornell.edu
These authors contributed equally: Tingxin Li, Shengwei Jiang, Lizhong Li.

**The evolution of a Landau Fermi liquid into a nonmagnetic Mott insulator with increasing electronic interactions is one of the most puzzling quantum phase transitions in physics [1-6]. The vicinity of the transition is believed to host exotic states of matter such as quantum spin liquids [4-7], exciton condensates [8] and unconventional superconductivity [1]. Semiconductor moiré materials realize a highly controllable Hubbard model simulator on a triangular lattice [9-22], providing a unique opportunity to drive a metal-insulator transition (MIT) via continuous tuning of the electronic interactions. Here, by electrically tuning the effective interaction strength in $MoTe_2/WSe_2$ moiré superlattices, we observe a continuous MIT at a fixed filling of one electron per unit cell. The existence of quantum criticality is supported by the scaling behavior of the resistance, a continuously vanishing charge-gap as the critical point is approached from the insulating side, and a diverging quasiparticle effective mass from the metallic side. We also observe a smooth evolution of the low-temperature magnetic susceptibility across the MIT and find no evidence of long-range magnetic order down to ~ 5% of the Curie-Weiss temperature. The results signal an abundance of low-energy spinful excitations on the insulating side that is further corroborated by the presence of the Pomeranchuk effect on the metallic side. Our results are consistent with the universal critical theory of a continuous MIT from a Landau Fermi liquid to a nonmagnetic Mott insulator in two dimensions [4,23].**

The interaction induced localization of electrons – the Mott transition – is expected to occur in the half-filled Hubbard model [1-3,24,25]. The ground state is a metal with a sharply defined electronic Fermi surface when the kinetic energy of the electrons – characterized by the bandwidth $W$ – far exceeds their interaction energy – characterized by the on-site Coulomb repulsion $U$. Conversely, when $U \gg W$, the ground state is an electrical insulator with a charge-gap. The system undergoes a MIT when $U$ and $W$ become comparable. Although this picture is widely accepted from the seminal works of Mott and Hubbard, the nature of the transition remains poorly understood. In most materials, the transitions are driven first-order and often accompanied by simultaneous magnetic, structural or other forms of ordering [1,3]. Continuous MIT, which exhibits no symmetry breaking, an abrupt disappearance of an entire electronic Fermi surface and the simultaneous opening of a charge-gap across a quantum critical point, remains one of the outstanding problems in condensed matter physics [1-6,26-28]. Despite the extensive theoretical studies on the topic [1-8, 23-29], experimental candidates remain scarce [1].

Continuous Mott transitions are generally favored by geometric frustration and reduced dimensionality, where strong quantum fluctuations can weaken or even quench different types of order [3-4,28-31]. Moiré heterostructures of two-dimensional (2D) semiconducting transition metal dichalcogenides (TMDs), which are believed to realize a triangular lattice Hubbard model [9,19,20], provide an ideal testbed of the Mott transition [21,22]. The system is highly controllable – allowing independent tuning of both the filling factor and the effective interaction strength ($U/W$). In particular, the electron density can be continuously tuned by gating in a field-effect device structure [10-16]. The effective interaction strength can be tuned, in principle, by varying the twist angle between the TMD layers [9,20], which determines the moiré period and thus the bandwidth. Here we demonstrate continuous tuning of $U/W$ by an out-of-plane electric field. The electric field varies the potential difference between the two TMD layers and subsequently the moiré potential, which changes the size of the localized Wannier function and the bandwidth predominantly. We investigate the electrical transport and magnetic properties of the system at fixed half-band filling as a function of effective interaction.

We choose angle-aligned $MoTe_2/WSe_2$ heterobilayers with hole doping. The two TMD materials have ~ 7% lattice mismatch. At zero twist angle they form a triangular moiré superlattice with period of ~ 5 nm (Fig. 1a), which corresponds to a moiré density of ~ $5\times10^{12}$ cm$^{-2}$. In each TMD monolayer, the band edges are located at the K/K' points of the Brillouin zone with double spin-valley degeneracy. The electronic band structures of relaxed zero-degree-aligned $MoTe_2/WSe_2$ heterobilayers are characterized from density functional theory (DFT) (Methods). They have type-I band alignment with valence band offset of ~ 220 meV (both conduction and valence band edges are from $MoTe_2$). Figure 1d illustrates the first two hole moiré bands under two values of out-of-plane displacement field, $D$. The field is along the direction that reduces the valence band offset. The valley degeneracy is lifted by valley pseudospin-orbit coupling in small-period moiré structures. The displacement field shows a strong effect on band dispersions. The bandwidth of the first (valley-split) moiré bands increases rapidly with displacement field for sufficiently large fields (Fig. 1e), indicating the feasibility of bandwidth-tuned MITs. The large lattice mismatch of the two materials also has several practical advantages. The heterostructures are less prone to effects of disorder from angle alignment inhomogeneities since the moiré period is not sensitive to twist angle near zero degree. The large moiré density, or equivalently, doping density at half filling compared to the disorder density (~ $10^{11}$ cm$^{-2}$, Methods) favors purely interaction-driven MITs. Finally, the large doping density facilitates the formation of good electrical contacts for transport measurements at low temperatures.

We fabricate dual-gate field-effect devices of $MoTe_2/WSe_2$ heterobilayers with hexagonal boron nitride (hBN) gate dielectrics and graphite gate electrodes (Fig. 1a, b). The typical hBN thickness is about 5 and 20-30 nm, respectively, in the top and bottom gates. We pattern the devices into Hall bar geometry and measure the four-point sheet resistance down to 300 mK (see Methods for device fabrication and measurements). Figure 1c shows square resistance $R_\square$ of Device 1 at 300 mK as a function of two gate voltages. It can be converted to resistance as a function of filling factor $f$ and applied out-of-plane electric field $E$ using the known device geometry (Extended Data Fig. 1). The two prominent resistance features correspond to $f = 1$ and 2, respectively, where $f = 1$ denotes one hole per moiré cell, that is, half filling of the first hole moiré band. Resistance increases substantially at $f < 1$. The insulating states at $f = 1$ and 2 are the known Mott and (trivial) band insulating states, respectively [10-16]. At large applied fields they both turn metallic. The MIT at $f = 2$ occurs at a smaller field because of the increasing bandwidth with fillings [14]. It induces no observable effects on the Mott insulating state. The result is presented in Extended Data Fig. 2. The applied field here closes the gap between the first and second moiré bands and induces a transition from a band-insulator to a compensated semi-metal. Such a transition occurs through a mechanism that is distinct from the Mott transition at $f = 1$.

Below we focus on the interaction (bandwidth)-driven MIT at $f = 1$. The sheet resistance or conductance ($G_\square = 1/R_\square$) is very sensitive to applied electric field near the transition. At the lowest temperature (285 mK), it changes by more than 4 orders of magnitude within a narrow range of the critical field $E_C \approx 1.304$ V/nm (Fig. 1f). No hysteresis is observed for different sweeping directions of the field.

Figure 2a illustrates the temperature dependence of resistance up to 70 K at representative electric fields. They show two types of behaviors. Below the critical field, the resistance increases as temperature $T$ approaches zero. This is characteristic of an insulator. The resistance follows a thermal activation dependence (Extended Data Fig. 3). We extract the activation gap $\Delta$ for charge transport in Fig. 2b. The gap size decreases monotonically from tens of meV to a few meV as $E_C$ is approached from below. It follows a power-law dependence $\Delta \propto |E - E_C|^{\nu z}$ with exponent $\nu z \approx 0.60 \pm 0.05$ (inset).

Above the critical electric field, the resistance follows a $T^2$-dependence at low temperatures over a temperature range up to ~ 10 K. This is characteristic of a Landau Fermi liquid with electron-electron umklapp scattering. We fit the low-temperature resistance with $R_\square = R_0 + AT^2$ (Extended Data Fig. 4), where $R_0$ denotes the residual resistance and $A^{1/2}$ is proportional to the quasiparticle effective mass $m^*$ according to Kadowaki-Woods scaling [32]. In Fig. 2c we plot the electric field dependence of $A^{1/2}$. The data is well described by a power-law divergence $A^{1/2} \propto m^* \propto |E - E_C|^{-1.4\pm 0.1}$ as $E$ approaches $E_C$ from above. The result suggests the entire electronic Fermi surface contributing to transport, with $m^*$ diverging at $E_C$ due to the effects of quantum fluctuations near the MIT [2-4,23,24,29].

At higher temperatures, the resistance deviates from the $T^2$-dependence and reaches a maximum at temperature $T^*$, above which the resistance decreases with increasing temperature. Upon approaching the MIT, the value of $T^*$ decreases (dashed line in Fig. 3b). The square resistance exceeds the Mott-Ioffe-Regel limit ($\frac{h}{e^2}$ with $h$ and $e$ denoting, respectively, the Planck's constant and the elementary charge). It corresponds to a mean free path smaller than the moiré period and is suggestive of 'bad' metallic behavior [33].

Next we demonstrate quantum critical scaling collapse of the resistance curves near the MIT. We first identify the precise value of the critical field following the standard procedure that a simple power-law dependence of $R_\square(T)$ is expected only at the critical point (Extended Data Fig. 5). We then normalize $R_\square(T)$ by resistance at the critical field $R_C(T)$. All resistance curves near the MIT collapse onto two branches after the temperatures are scaled by field-dependent $T_0$'s (Fig. 3a). The top and bottom branches represent the insulating and metallic transport behaviors, respectively; they display reflection symmetry about $R_\square/R_C = 1$ in the log-log plot. The scaling parameter $T_0$ continuously vanishes as the critical field is approached from both sides (Fig. 2b). Similar to the charge-gap, $T_0$ follows a power-law dependence $T_0 \propto |E - E_C|^{\nu z}$ with practically identical exponent $\nu z \approx 0.70 \pm 0.05$ for both sides.

We also show the field-temperature phase diagram for $\left|\log \frac{R_\square}{R_C}\right|$ in Fig. 3b. It reveals the 'fan-shape' structure that is widely observed for quantum criticality [29,30,31]. The Widom line is close to the vertical blue line stemmed from the critical field (Methods). And $T_0$ (dashed line) sets the scale for the finite temperature crossovers near the MIT, that is, the boundary of the quantum critical region [27]. The $T^*$ line is on the metallic side.

Since the ground and low-energy excited states of the Mott insulator are determined by magnetic interactions, we examine the magnetic properties near the critical point. A magnetic field parallel to the 2D plane couples weakly to spins because of the strong Ising spin-orbit interaction in TMDs [34] (Extended Data Fig. 6). We characterize the magnetization of holes in TMD moiré heterostructures under an out-of-plane magnetic field $B$ by magnetic circular dichroism (MCD), a method that has been demonstrated in a recent study [11] (Methods). Figure 4a illustrates the magnetic-field dependence of the magnetization for several electric fields at 1.6 K. The magnetization initially increases linearly with $B$ and then saturates at an electric-field dependent $B^*$ (symbols). The values of $B^*$ agree well with the saturation fields of magnetoresistance measured on the metallic side (Fig. 4c). We determine the saturation field as the field at which the transport behavior crosses over from metallic to insulating as shown in the inset for one electric-field example.

Next we obtain the magnetic susceptibility $\chi$ from the slope of MCD around $B = 0$. Figure 4b shows the temperature dependence of $\chi^{-1}$ at varying electric fields. On the metallic side, $\chi$ saturates at low temperatures. Above a temperature $\sim T^*$ (marked by arrows), the susceptibility is well described by the Curie-Weiss dependence $\chi^{-1} \propto T - \theta$ (dashed lines) with a negative Weiss constant $\theta$ ($\sim$ 30-40 K), which reflects the antiferromagnetic interaction between the local moments in the Hubbard model [2]. The magnetic susceptibility shows a smooth dependence on temperature for all electric fields (Fig. 4b) and on electric field across the MIT down to 1.6 K (Fig. 4d).

The magnetic properties are well correlated with the transport properties in Fig. 2a. At low temperatures, the system is a Landau Fermi liquid on the metallic side; $\chi$ is given by the Pauli susceptibility of the heavy fermions near the electronic Fermi surface [3]. Above $T^*$, the system enters an incoherent regime; local moments emerge; and the susceptibility follows the Curie-Weiss dependence. This behavior is reminiscent of the Pomeranchuk effect observed in helium-3 (also see Extended Data Fig. 7), where the increasing localization and formation of local moments lead to an increase in spin entropy with increasing temperature [23,31,35]. The coherent quasiparticles can also be destroyed when the Zeeman energy at magnetic saturation overcomes the renormalized bandwidth [3] ($g\mu_B B^* \gtrsim W^*$, Methods). This picture is consistent with the magnetoresistance data in Fig. 4c and is further supported by the good agreement between $g\mu_B B^*$ and the thermal excitation energy ($k_B T^* \sim W^*$) in Fig. 3b. Here $g$, $\mu_B$ and $k_B$ denote the hole Lande g-factor ($g \approx 11$ in TMDs [36]), the Bohr magneton, and the Boltzmann constant, respectively. In contrast to most 2D electron systems [28,37], the hole Zeeman energy in TMD moiré superlattices is substantially larger than the cyclotron energy [38] because of the large g-factor and the heavy band mass that is further enhanced by the moiré flat bands.

Near the MIT, the magnetic interaction $J$ sets the smallest energy scale of the system since both $U$ and $W$ are tens of meV. For $T \ll \theta$, we can determine $J \sim g\mu_B B^* \sim 3$ meV from the insulating side (last panel of Fig. 4a), the value of which is in good agreement with the extracted Weiss constant $k_B\theta \sim 3$ meV. The thermal excitation energy at the lowest measurement temperature (1.6 K and 300 mK for magnetic and transport properties, respectively) is well below this scale. Therefore, the smooth temperature dependence of $\chi$ without any sign of spin-gap for all electric fields (Fig. 4b) and the smooth evolution of $\chi$ across the MIT (Fig. 4d) show the absence of long-range magnetic order on both sides. These observations point to a MIT from a Fermi liquid to a nonmagnetic (or 120-degree Néel below 1.6 K) Mott insulator with extensive spin entropy at finite temperatures. This is expected for a frustrated lattice [4,23,30,31] and further corroborated by the Pomeranchuk effect. Moreover, since $m^*$ diverges from the metallic side, the smooth evolution of $\chi$ across the MIT implies a diverging Landau-parameter, $F_0^a$, and similarly, a diverging $F_0^s$ as the compressibility must vanish at the MIT [4].

In conclusion, we have demonstrated a continuous Mott transition in MoTe$_2$/WSe$_2$ moiré superlattices down to 300 mK and performed scaling analyses on major physical quantities near the quantum critical point. In contrast to MITs induced by tuning the doping density in conventional 2D electron systems [3,28,37], the carrier density here is fixed at half-band filling. The MIT is induced by varying an out-of-plane electric field that modifies the moiré potential depth and thus $U/W$. Our results point to a clear example of a continuous MIT across which the entire electronic Fermi surface disappears abruptly. In addition, because the half-band filling density is nearly two orders of magnitude higher than the disorder density, disorder only plays a perturbative role in the observed interaction-driven MIT. This is in stark contrast to the density-tuned MITs; there both the interaction and disorder effects are important (Mott-Anderson transitions [3]). Our primary observations near the MIT are consistent with the continuous Mott transition from a Landau Fermi liquid to a nonmagnetic Mott insulator [4,23], including a continuously vanishing charge-gap, a diverging effective mass, a constant spin susceptibility across the MIT, and the Pomeranchuk effect. Future investigations of the transport and magnetic properties near the transitions, particularly at lower temperatures, may reveal new exotic states of matter such as quantum spin liquids.

**Methods**
**Device fabrication.** We fabricated angle-aligned $MoTe_2/WSe_2$ devices using the layer-by-layer dry transfer method [39]. The constituent atomically thin flakes were exfoliated from bulk crystals onto Si substrates with a 285-nm oxide layer and picked up by a polycarbonate (PC) stamp in desired sequence. We first released a stack of graphite/hBN layers onto a $Si/SiO_2$ substrate as the bottom gate. Platinum (Pt) contacts were patterned into Hall bar geometry on hBN by e-beam lithography and metallization. The second stack consisting of a $MoTe_2/WSe_2$ bilayer and an hBN/graphite top gate was released onto the pre-patterned Pt electrodes. Figure 1a and 1b are a schematic side view and an optical micrograph of a typical device, respectively. We used angle-resolved optical second-harmonic generation (SHG) spectroscopy [11] to determine the crystal orientations of $WSe_2$ and $MoTe_2$ monolayers and the twist angle between them in the bilayer. The devices investigated in this study are zero-degree aligned within ± 0.5°. The uncertainty is mainly limited by that of the SHG measurements. We used relatively thin hBN layers (~ 5 nm) in the top gate. We found that in general thinner hBN layers can sustain a larger breakdown electric field.

Several steps were taken to produce high-quality $MoTe_2/WSe_2$ devices. First we kept the Pt contacts thin (~ 5 nm) to reduce strain on the device. Second, we removed the polymer residues on the bottom gate from Pt electrode fabrication using an atomic force microscope (AFM) in contact mode. Finally, we handled atomically thin $MoTe_2$ flakes inside a nitrogen-filled glovebox with oxygen and water levels below 1 part per million (ppm) to minimize degradation of $MoTe_2$. The device spatial inhomogeneity was evaluated by probing the transport properties using various source-drain pairs of a multi-terminal device (Extended Data Fig. 8). We have studied a total of 4 devices. All of them show interaction-tuned MIT at $f = 1$ and 2 but they have varying degrees of sample inhomogeneity. More homogeneous devices allow scaling analysis closer to the quantum critical point. The results for Device 1, which has the best sample homogeneity, are presented in the main text. The results of Device 2 are shown in Extended Data Fig. 9.

**Electrical measurements.** Electrical transport measurements were performed in a closed-cycle $^4$He cryostat (Oxford TeslatronPT) equipped with a superconducting magnet and a $^3$He insert (base temperature ~ 300 mK). Standard low-frequency (< 23 Hz) lock-in techniques were used to measure the sample resistance under a bias voltage of 1-2 mV. Both the voltage drop at the probe electrode pairs and the source-drain current were recorded. Voltage pre-amplifiers with large input impedance (100 MΩ) were used to measure sample resistance up to ~ 10 MΩ.

**MCD measurements.** Magnetic circular dichroism (MCD) was performed in an Attocube closed-cycle optical cryostat (attoDry2100) down to 1.6 K and under an out-of-plane magnetic field of up to 9 T. The optical beam was focused onto the sample using an optical microscope objective (0.8 numerical aperture); the beam diameter was ~ 1 μm on the device. The reflected light was collected by the same objective and directed to detectors.

We first characterized the MCD spectrum using a broadband tungsten-halogen lamp. A combination of a linear polarizer and an achromatic quarter-wave plate was used to generate circularly polarized light. The incident power on the device was kept below 1 nW. The left- or right-handed light reflected from the sample was spectrally resolved by a spectrometer coupled to a liquid nitrogen-cooled charge coupled device (CCD). The MCD at a given photon energy is defined as MCD $\equiv (I_L - I_R)/(I_L + I_R)$, where $I_L$ and $I_R$ are the intensity of the left and right circularly polarized light, respectively. A sample MCD spectrum as a function of out-of-plane electric field is shown in Extended Data Fig. 10. The MCD is strongly enhanced near the fundamental exciton resonance of WSe$_2$; the electric field has a negligible effect on the resonance energy for electric fields near $E_C$. This allows us to probe the MCD response at a fixed wavelength 747.4 nm (Fig. 4). The optical excitation was provided by a tunable, continuous-wave Ti-Sapphire laser (M Squared SOLSTIS system). We limited the incident light power to about 300 nW on the sample to minimize the heating effects. We modulated the light helicity by a photoelastic modulator at 50.1 kHz and detected the reflected light by a photodiode. The MCD signal is defined as the ratio of the modulated signal (measured by a lock-in amplifier) to the total reflected light power (measured by a DC voltmeter).

**Band structure calculations.** Density functional calculations were performed using Perdew-Burke-Ernzerhof generalized gradient approximation [40] with the van der Waals correction incorporated by DFT-D3 method with Becke-Jonson damping [41] as implemented in the Vienna Ab initio Simulation Package [42]. Pseudopotentials were used to describe the electron-ion interactions. We first constructed the zero-degree-aligned MoTe$_2$/WSe$_2$ heterobilayer with vacuum spacing larger than 20 Å to avoid artificial interaction between the periodic images along the out-of-plane direction. The structure relaxation was performed with force on each atom less than 0.01 eV/Å. We used Gamma-point sampling for structure relaxation and self-consistent calculation.

**Effects of disorders.** Because of sample inhomogeneity, the scaling collapse of resistance curves in Fig. 3 fails for $|E - E_C| < 0.002$ V/nm. We therefore cannot reliably obtain $T_0 \lesssim 20$ K from scaling. The corresponding energy scale is, however, many times smaller than $U \sim W \sim 70$ meV near $E_C$ (estimated from the Coulomb repulsion energy corresponding to the moiré period) so that we are safely in the low-temperature limit. We can also estimate the disorder density in Device 1 as $\sim 2\varepsilon_{hBN}\varepsilon_0 \times (0.002 \text{V/nm}) \sim 10^{11}$ cm$^{-2}$. Here $\varepsilon_{hBN} \approx 3$ and $\varepsilon_0$ are the out-of-plane dielectric constant of hBN and the vacuum permittivity, respectively. The disorder density is about two orders of magnitude smaller than the moiré density.

**Widom line.** We adopted the notion of a generalized Widom line [29] to separate regions of metallic and insulating behavior in the experimental electric field-temperature phase diagram (Fig. 3b). In the scaling analysis of resistance (Extended Data Fig. 5), we find a simple power-law dependence of $R_\square(T)$ at one particular electric field, which are identified as the 'separatrix' $R_C(T)$ and the critical field $E_C$. The existence of $R_C(T)$ at a constant electric field implies that the Widom line is close to a vertical line stemmed from $E_C$ in the phase diagram. We verified this by finding the inflection points of logarithmic resistance as a function of electric field ($\frac{d^2(\log R_\square)}{dE^2} = 0$) at different temperatures. The $\log R_\square$ inflection point line has been shown to well represent the Widom line. Extended Data Fig. 5 shows that the inflection points are nearly temperature independent, i.e. a vertical Widom line from $E_C$.

**Estimate of the renormalized bandwidth of the heavy Fermi liquid near MIT.** Based on the data in Fig. 2c, $m^*$ is enhanced from the single-particle moiré band mass by at least two orders of magnitude near $E_C$. We therefore expect the renormalized bandwidth $W^*$ of the heavy Fermi liquid to be greatly reduced from the single-particle bandwidth $W \sim U \sim 70$ meV near $E_C$. Because $m^*$ is only a property of the Fermi surface, however, we cannot directly relate $W^*$ to $1/m^*$. Nevertheless, the significant reduction in the bandwidth near $E_C$ leads to $W^* \sim g\mu_B B^* \sim k_B T^*$ as shown in Fig. 3b.

**Quantum oscillations in the magnetoresistance.** The magnetoresistance of the MoTe$_2$/WSe$_2$ device in Fig. 4c shows Shubnikov-de Hass-like oscillations in addition to the metal-insulator transition. These oscillations are associated with formation of Landau levels in the top graphite gate under an out-of-plane magnetic field. They vanish above ~30 K, which is in good agreement with the reported temperature range for Shubnikov-de Hass oscillations in graphite [43]. They also vanish when we replace graphite by 2D metal TaSe$_2$ in the top gate (Extended Data Fig. 11). The coupling between the closely spaced MoTe$_2$/WSe$_2$ sample and the top graphite gate (~ 5 nm) is presumably capacitive, i.e. through screening [44]. The energy gap for charge excitations in a 2D insulator is sensitive to its dielectric surroundings [45]. When Landau levels are developed in the nearby graphite gate under a magnetic field, the oscillations in graphite's density of states induce oscillatory changes in the effective dielectric function that the 2D insulator experiences, and consequently, oscillations in the charge-gap and the in-gap resistance (through thermal activation) of the sample. The oscillation amplitude is the largest for in-gap electrical transport (Extended Data Fig. 11).

**Acknowledgements**
We thank Senthil Todadri, Eun-Ah Kim, Vladimir Dobrosavljevic, Louk Rademaker and Allan H. MacDonald for fruitful discussions.

# Figures

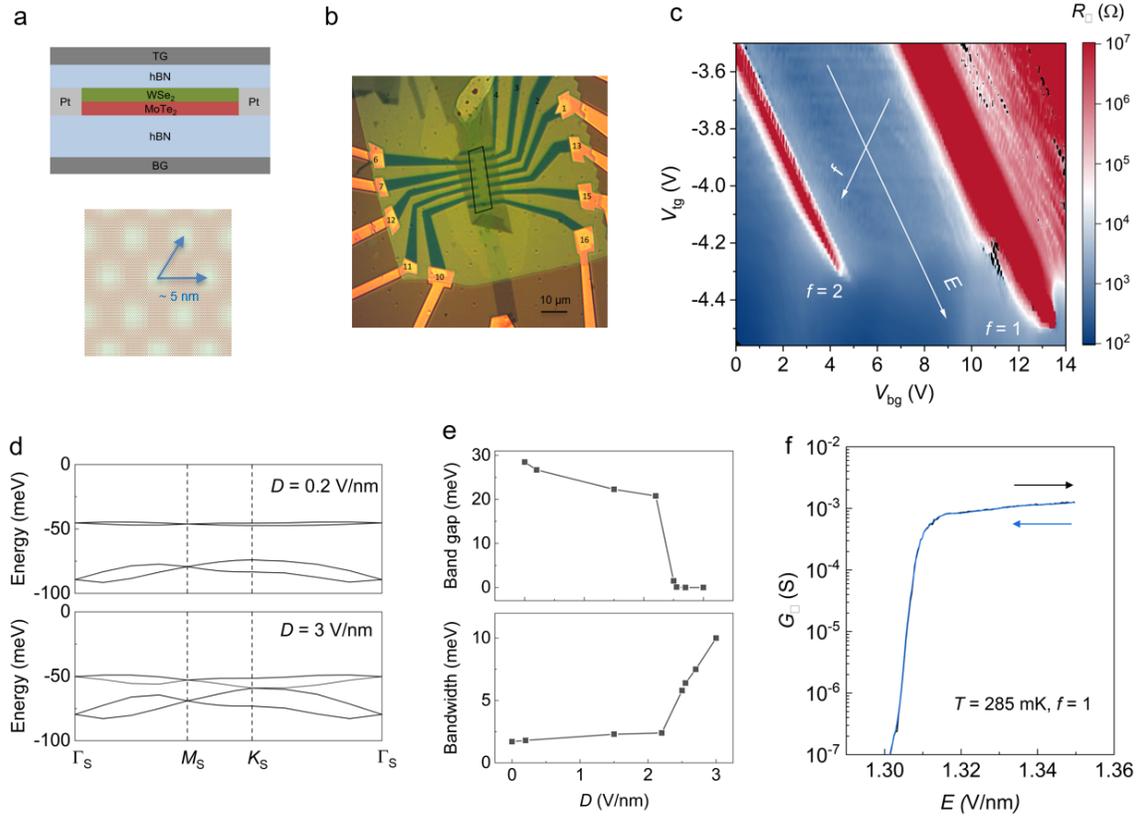

**Figure 1 | Bandwidth-tuned metal-insulator transition. a,** Top: Schematic side view of MoTe$_2$/WSe$_2$ moiré heterobilayer devices with hBN/graphite gates and Pt contact electrodes. The two gates allow independent tuning of filling factor and out-of-plane electric field. Bottom: The moiré superlattice structure with 5 nm moiré period. **b,** Optical microscope image of multi-terminal Hall bar Device 1. The conduction channel of the MoTe$_2$/WSe$_2$ moiré heterobilayer is outlined by black lines. The scale bar represents 10 $\mu$m. **c,** Square resistance of Device 1 at 300 mK in logarithmic scale as a function of top and bottom gate voltages. The gate voltages relate to the hole filling factor *f* and electric field *E* (field direction is up in **a**). Electric field-induced MIT is observed at $f = 1$ and 2. **d,** Electronic band structure (first two hole moiré bands) of zero-degree-aligned MoTe$_2$/WSe$_2$ heterobilayers from DFT under displacement field $D = 0.2$ V/nm (top) and 3.0 V/nm (bottom). **e,** Band gap between the first and second moiré bands (top) and bandwidth of the first moiré bands (bottom) as a function of displacement field from DFT. **f,** Square conductance at $f = 1$ as a function of electric field near the MIT at 285 mK. No hysteresis is observed under forward and backward scans of the electric field.

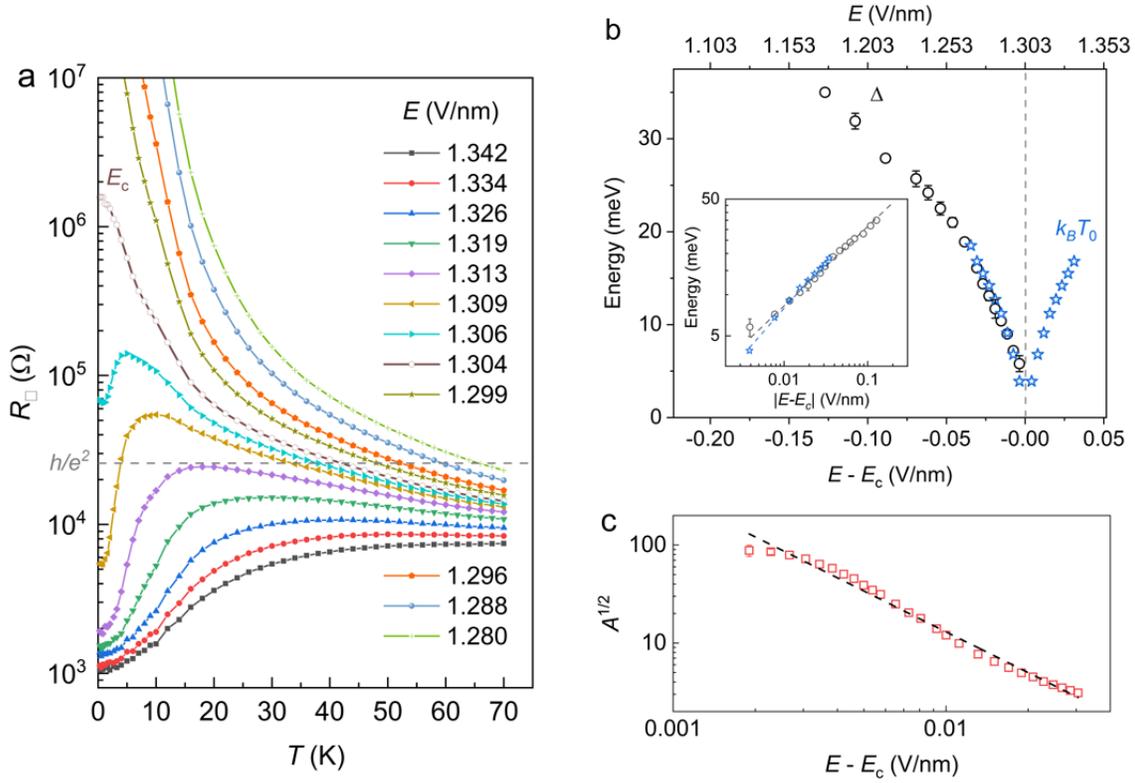

**Figure 2 | Continuous Mott transition. a,** Temperature dependence of square resistance (symbols) at $f = 1$ under varying electric fields. The lines are guide to the eye. The resistance at $E_C$ = 1.304 V/nm (open symbols) follows a power-law dependence. The horizontal dashed line marks the resistance quantum. **b,** Continuously vanishing charge-gap $\Delta$ and temperature scaling parameter $T_0$ (multiplied by $k_B$) as the critical field is approached. Both follow a power-law dependence on $|E - E_C|$ with nearly identical exponents (inset). **c,** Electric-field dependence of $A^{1/2}$ in a log-log plot, where $A$ is the fitting parameter for the low-temperature square resistance ($R_\square = R_0 + AT^2$) and the error bars are the fitting uncertainty. The dashed line is a power-law fit $A^{1/2} \propto |E - E_C|^{-1.4 \pm 0.1}$.

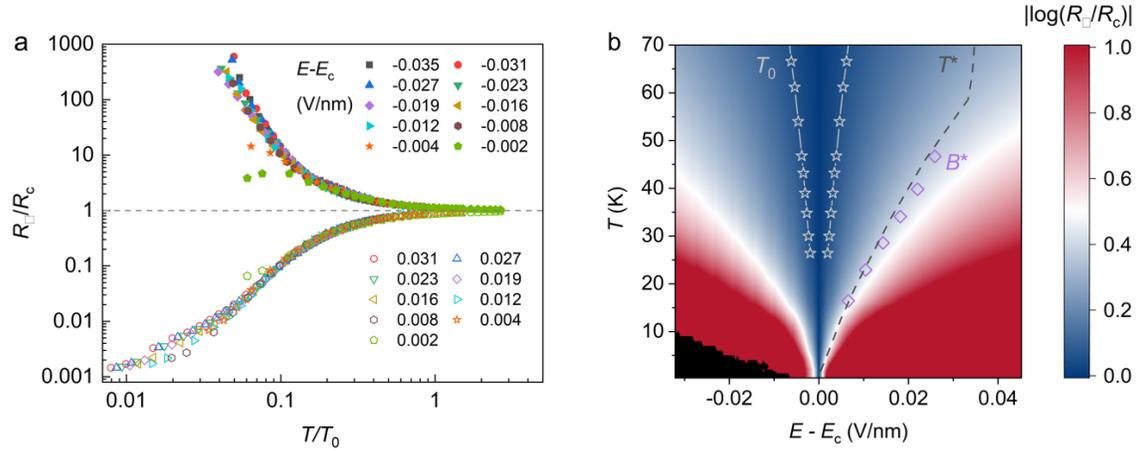

**Figure 3 | Quantum critical scaling. a,** Temperature-dependent square resistance curves near the MIT collapse onto two branches. The resistance curves are scaled by that at the critical field $R_C(T)$; the temperatures are scaled by field-dependent $T_0$'s. **b,** Electric field-temperature phase diagram for $\left|\log\frac{R_\square}{R_C}\right|$. White symbols are the temperature scaling parameter $T_0$'s. The lines are guide to the eye; they set the quantum critical regions. The dashed line corresponds to the crossover temperature $T^*$. Purple symbols are the equivalent temperature for the Zeeman energy at the saturation fields of magnetoresistance.

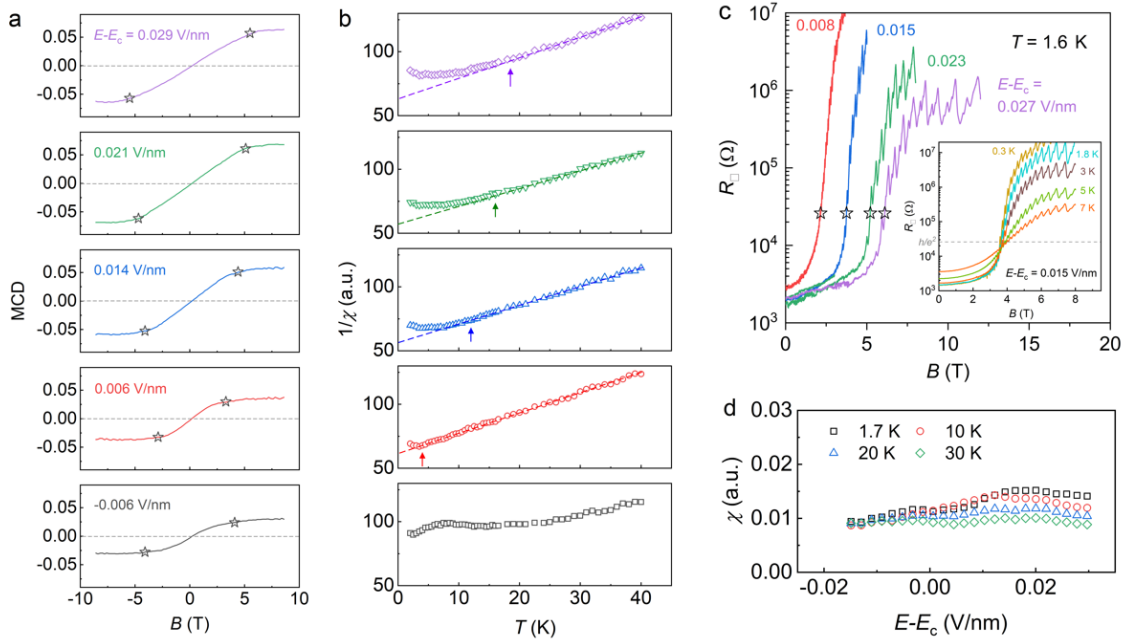

**Figure 4 | Magnetic properties near the Mott transition. a,** MCD as a function of out-of-plane magnetic field $B$ under varying electric fields at 1.6 K. It increases linearly with $B$ for small fields and saturates at $B^*$ (symbols). **b,** Temperature dependence of the inverse magnetic susceptibility $\chi^{-1}$ determined from the data in **a**. On the metallic side, $\chi$ saturates at the lowest temperatures; it follows the Curie-Weiss dependence (dashed lines) above the crossover temperatures (denoted by arrows) from a Fermi liquid to an incoherent metal. Non-monotonic temperature dependence is observed for $\chi$ on the insulating side. **c,** Magnetoresistance at varying electric fields above the critical electric field (1.6 K). Inset: magnetoresistance at varying temperatures for one of the electric fields; a magnetic field-induced MIT is observed. The crossover magnetic-field value is used to estimate the saturation field $B^*$ (symbols) in the main panel. It agrees well with the value from the MCD measurement in **a**. The quantum oscillations observed in magnetoresistance are associated with quantum oscillations in the top graphite gate (see Methods for details). **d,** Smooth evolution of magnetic susceptibility at varying temperatures through the quantum critical point, demonstrating the absence of magnetic phase transition.

**Extended Data Figures**

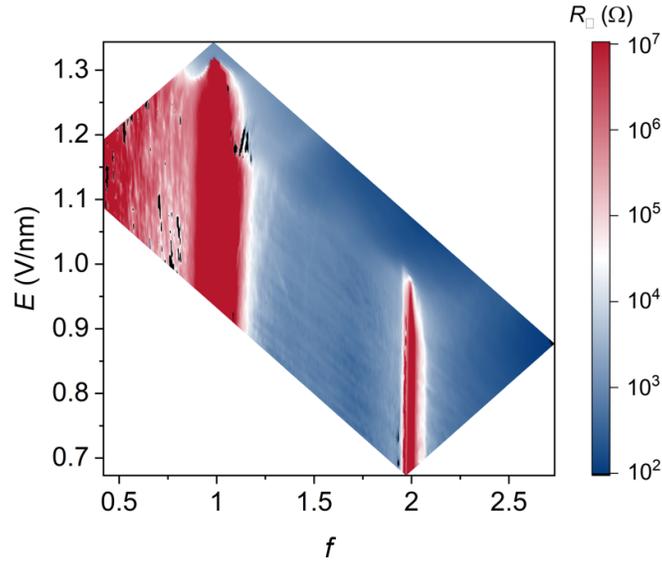

**Extended Data Figure 1 | Square resistance versus electric field and filling factor.** 2D map of electric field and filling factor dependence of the square resistance (in log scale) at 300 mK converted from the data in Fig. 1c. Electric field-induced MITs are seen at both $f = 1$ and $f = 2$.

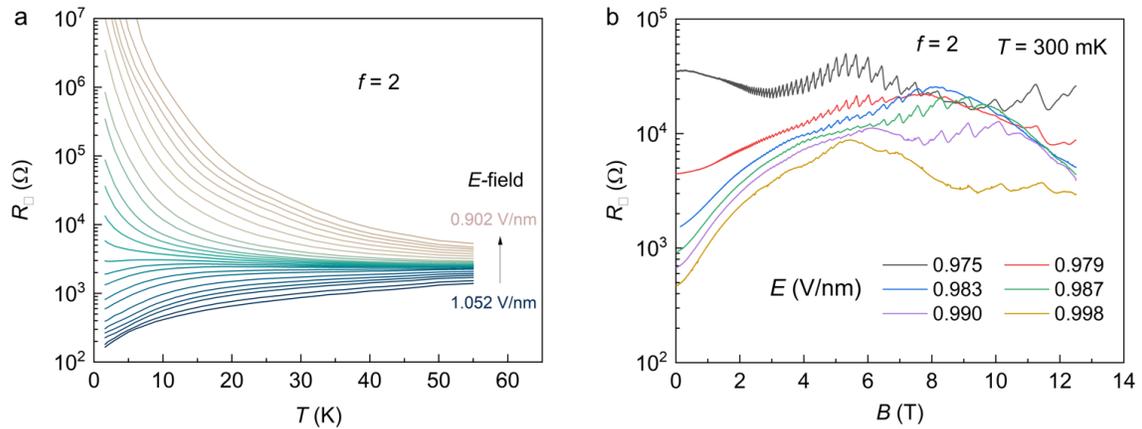

**Extended Data Figure 2 | Metal-insulator transition at $f = 2$. a,** Temperature dependence of the square resistance at varying electric fields at $f = 2$. A metal-insulator transition is observed near 0.978 V/nm. Compared to the $f = 1$ MIT, no clear Landau Fermi liquid behavior and no Pomeranchuk effect are observed on the metallic side in the studied temperature range. **b,** Magnetoresistance at varying electric fields at 300 mK. The behavior is diametrically different compared to that at $f = 1$. In particular, no magnetic field-induced metal-insulator transition is observed.

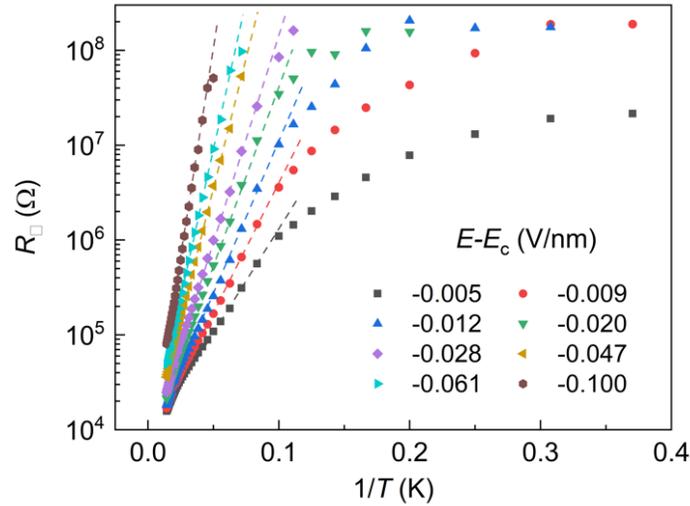

**Extended Data Figure 3 | Extraction of activation gap at $f = 1$.** Arrhenius plot of the square resistance (log scale) at varying electric fields. Thermal activation behavior is seen at high temperatures from which the activation gap can be extracted from the slopes of the linear fits (dashed lines).

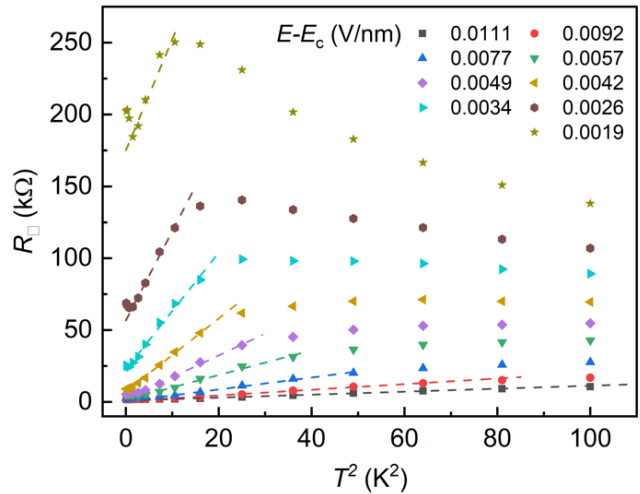

**Extended Data Figure 4 | Landau Fermi liquid behavior at low temperatures.** Dependence of the square resistance on temperature squared at varying electric fields. Landau Fermi liquid behavior with linear dependence is seen at low temperatures. The slope $A \propto (m^*)^2$ increases substantially near the critical electric field.

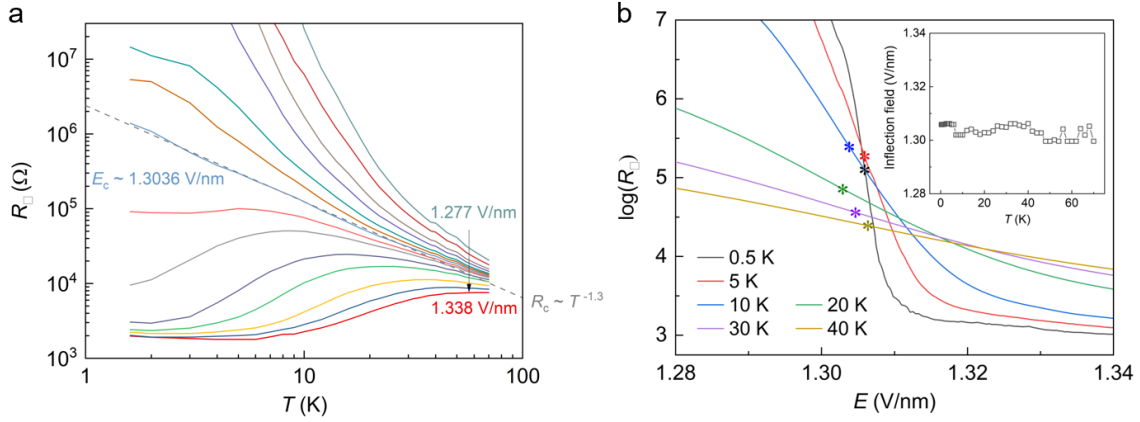

**Extended Data Figure 5 | Resistance scaling at $f = 1$ near critical point. a,** Dependence of square resistance on temperature at varying electric fields in log-log plot. A linear dependence is seen at the critical electric field, demonstrating the power-law temperature dependence. **b,** Electric field dependence of $\log R_\square$ at different temperatures. The inflection points are marked by the color symbols. The inset shows the temperature dependence of the electric field at the inflection point. The data shows that the Widom line is nearly a vertical line in Fig. 3b.

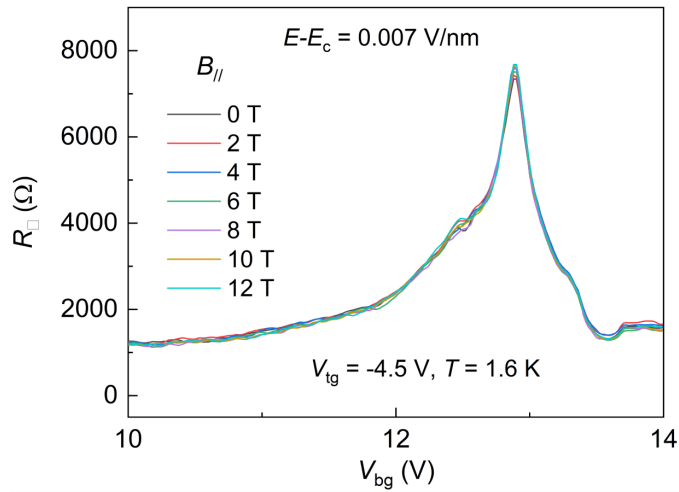

**Extended Data Figure 6 | Absence of in-plane magnetic field dependence.** Dependence of square resistance on the bottom gate voltage at varying in-plane magnetic fields. The bottom gate voltage primarily changes the filling factor $f$. The electric field is fixed at 0.007 V/nm (from $E_C$) near $f = 1$. No in-plane magnetic field dependence is observed due to the strong Ising spin-orbit coupling in monolayer TMDs.

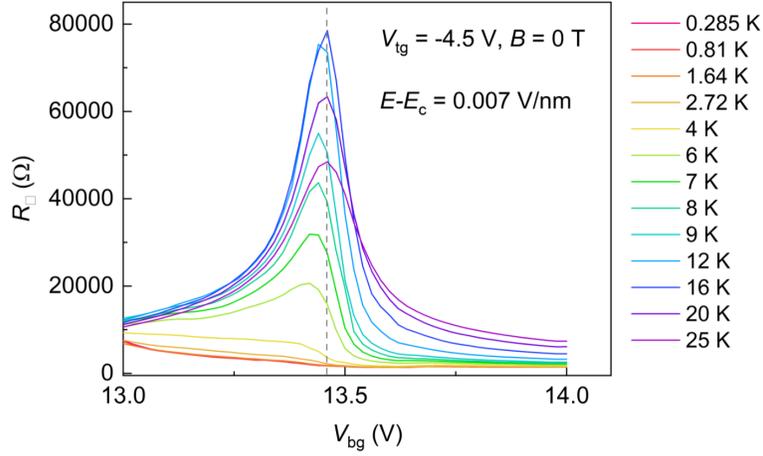

**Extended Data Figure 7 | Pomeranchuk effect at $f = 1$.** Dependence of square resistance on the bottom gate voltage at varying temperatures. The bottom gate voltage mainly changes the filling factor $f$. The electric field is fixed at 0.007 V/nm (from $E_C$) near $f = 1$. The $f = 1$ insulating state emerges with increasing temperature, demonstrating the Pomeranchuk effect. The result is fully consistent with the results presented in the main text, where the filling factor is kept constant at $f = 1$.

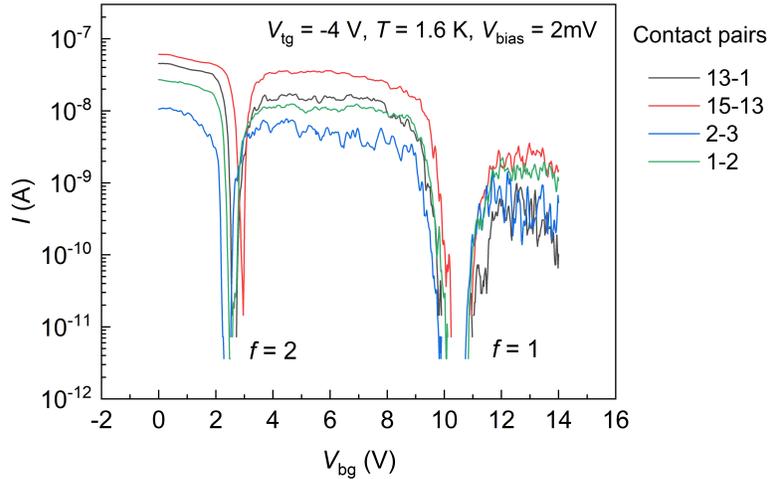

**Extended Data Figure 8 | Spatial homogeneity of Device 1.** Two-point current as a function of bottom gate voltage at fixed top gate voltage. The excitation bias voltage is 2 mV. The insulating states at $f = 1$ and $f = 2$ are seen at different source-drain pairs corresponding to the optical image in Fig. 1b. Some degrees of sample inhomogeneity are reflected by the slightly different positions of the insulating states. The two-point resistance also varies from pair to pair, reflecting the variation in contact/sample resistance.

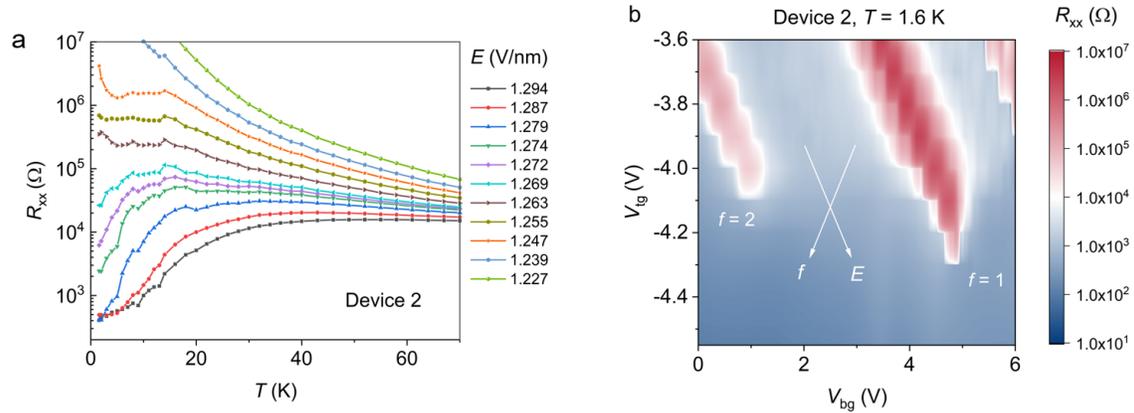

**Extended Data Figure 9 | Major results for Device 2. a,** Temperature dependence of the longitudinal resistance at $f = 1$ under varying electric fields. The critical electric field is near $E_C = 1.26$ V/nm. An MIT similar to Device 1 is observed. **b,** Longitudinal resistance at 1.6 K in logarithmic scale as a function of top and bottom gate voltages. The gate voltages relate to the hole filling factor $f$ and electric field $E$. Electric field-induced MIT is observed at $f = 1$ and 2. Compared to Device 1, there is a higher degree of spatial inhomogeneity in Device 2, which prevents reliable scaling analysis near the critical point.

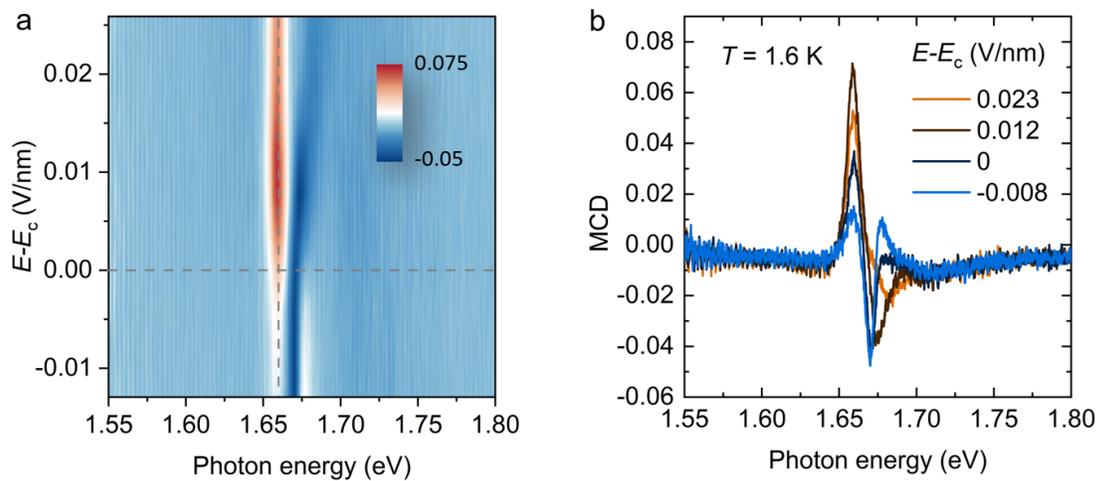

**Extended Data Figure 10 | MCD spectrum at 3 T perpendicular magnetic field. a,** Electric field dependence of the MCD spectrum near the WSe$_2$ exciton resonance. A clear resonance enhancement near 1.66 eV is observed. The vertical dashed line marks the photon energy of the probe laser beam used for the MCD measurements in Fig. 4 and the horizontal dashed line marks the MIT critical point. **b,** MCD spectra at selected electric fields illustrating the resonance enhancement near the exciton peak.

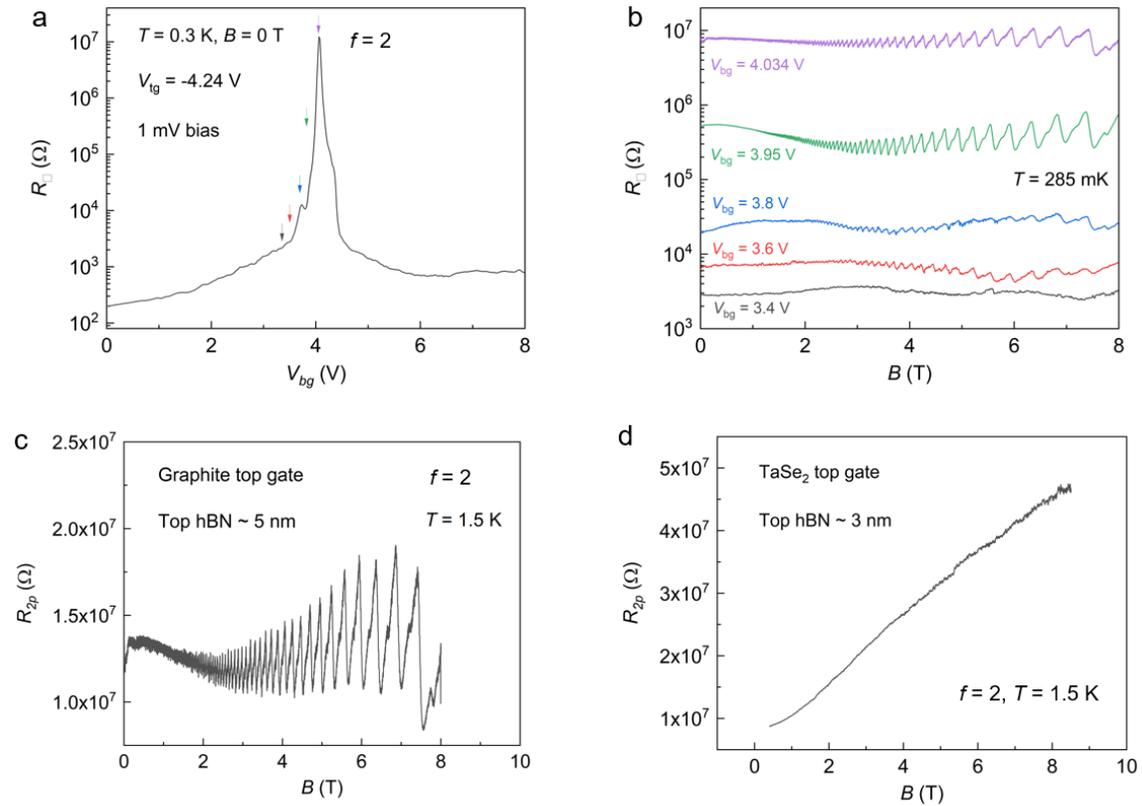

**Extended Data Figure 11 | Quantum oscillations in the insulating states. a,** Dependence of square resistance on the bottom gate voltage at 300 mK. The $f = 2$ insulating state is labeled. **b,** Magnetoresistance under a perpendicular magnetic field at selected bottom gate voltages marked by the arrows in **a**. Quantum oscillations due to the nearby graphite gate are observed near the insulating state. The oscillations disappear away from the $f = 2$ insulating state. **c,** Two-terminal magnetoresistance at the $f = 2$ insulating state with a nearby graphite gate that is ~ 5 nm separated from the sample. **d,** The same as **c** except the nearby graphite gate is replaced by a few-layer metallic $TaSe_2$ gate that is ~ 3 nm away from the sample. No quantum oscillations are developed in both the $TaSe_2$ gate and the sample under the studied magnetic field range. The results confirm that the quantum oscillations are originated from the high mobility graphite gate.